\documentclass[reprint,...]{revtex4-1}
\usepackage{url}
\usepackage{graphicx}

\usepackage{caption}
\usepackage{subcaption}
\usepackage{hyperref}

\begin{document}

\title{Using multiple-criteria methods to evaluate community partitions }

\author{ Cazabet Remy}\email{remy.cazabet@gmail.com}
\author{Chawuthai Rathachai}
\author{Takeda Hideaki}

\affiliation{National Institute of Informatics\\
2-1-2 Hitotsubashi, Chiyoda-ku,Tokyo,
Japan}



\date{\today}

\begin{abstract}
Community detection is one of the most studied problems on complex networks. Although hundreds of methods have been proposed so far, there is still no universally accepted formal definition of what is a good community. As a consequence, the problem of the evaluation and the comparison of the quality of the solutions produced by these algorithms is still an open question, despite constant progress on the topic. In this article, we investigate how using a multi-criteria evaluation can solve some of the existing problems of community evaluation, in particular the question of multiple equally-relevant solutions of different granularity. After exploring several approaches, we introduce a new quality function, called MDensity, and propose a method that can be related both to a widely used community detection metric, the Modularity, and to the Precision/Recall approach, ubiquitous in information retrieval.
\end{abstract}


\maketitle

\section{Introduction}
Community detection is one of the most studied problems on complex networks. Countless papers have been published on this topic, in particular during the last 15 years. However, most of this publication frenzy has been centered on the problem of proposing new methods. 

As the number of proposed methods increased, the problem of comparing them became more and more important. As a consequence, some of the most influential works on the subject propose methods to compare partitions between themselves. As the number of scalable methods increases, the question of which method to use becomes more and more important. 

In this paper, we will first review some of the existing methods to evaluate the quality of a partition. In a second section, we will discuss on what are good communities and good partitions, and will argue for the interest of using a two-criteria approach. Finally, we will go step by step from a two-criteria method directly inspired from information retrieval techniques to a more relevant method grounded on the Modularity.

\section{Evaluation Methods}
Several techniques have already been proposed to compare partitions --and therefore the algorithms that produced them-- between themselves. We can classify these methods in three families:
\begin{itemize}
    \item Single score metrics
    \item Evaluation on generated networks
    \item Evaluation on real networks with ground truth
\end{itemize}

We will review in the next sections these three types of evaluation methods

\subsection{Single score metrics}
Single score metrics are using quality functions associating a score to a given partition of a given network. To compare two partitions on a same network, one can simply compare their scores according to this quality function. Historically, the ancestor of community detection is the problem of graph partitioning --finding sets of nodes of predefined sizes such as the number of edges between them is minimum. In this problem, the quality function was simply the number of edges between communities. As this metric loses its significance as soon as we do not fix the size and number of the clusters to find, different metrics are used to evaluate the quality of community partitions. Several of these metrics are detailed in \cite{fortunato2010community}, while \cite{yang2012defining} compares some of them on real cases. However, the quality function that is by far the most popular is the Modularity. Initially introduced in \cite{girvan2002community}, this quality function is defined as the difference between the ratio of edges that fall inside communities and the expected value in a randomized version of the network. We will talk in details of the Modularity in the later sections of this article. It is now so popular that it is sometimes considered and used as a definition of communities. However, since the demonstration of its resolution limit \cite{fortunato2007resolution}, the usage of the sole Modularity to evaluate communities is discouraged. Adaptations of the Modularity called resolution-free methods have been proposed, notably in \cite{arenas2008analysis,reichardt2006statistical}, but have also been criticized. Surprise \cite{Aldecoa:2013fk} is another interesting function proposed recently, measuring how unlikely is a given partition compared to a null model.

Using a single score metric to evaluate communities has advantages: 
\begin{itemize}
    \item One can compare partitions not only on networks with a known solution but on any particular network of interest for him.
    \item For any network, a best solution can be unambiguously identified.
\end{itemize}

But also some drawbacks:
\begin{itemize}
    \item Using a quality function is using a fix definition of what is a community. However, this definition is arbitrary, as there is no consensus of the topic.
    \item On real networks, quality functions are likely to have only one maximal value. However, it is known that networks can have several meaningful levels of decomposition. Among several potential problems raised by this observation, a perfect partition at a suboptimal level might get a lower score than an imperfect decomposition at the optimal level.
\end{itemize}

\subsection{Evaluation on generated networks}
This approach consists in generating random networks with a well-defined community structure, known by construction, running several algorithms on these networks, and check how well the partitions found match the expected ones. From ad-hoc methods such as the one used in \cite{girvan2002community}, more advanced network generators have been proposed. The LFR benchmark \cite{lancichinetti2008benchmark} is the most widely used, it allows to tune several parameters such as the number of nodes, average degrees, distribution of the size of communities, and so on and so forth. 
To compare the solutions found to the reference, the most widely used function is the Normalized Mutual Information \cite{lancichinetti2009detecting}, but other approaches are possible, such as topological approaches \cite{orman2012comparative}, or cluster-analysis methods modified to take into account the network structure \cite{labatut2013generalized}.

Advantages:
\begin{itemize}
    \item Although proposing networks with a community structure requires a loose definition of what is a community, consensual results are easier to reach than in a quality function, where the definition is formal. It is easier to recognize a good community when we saw one that to give a universal definition.
    \item Variations on usual communities, such as hierarchical decomposition, fuzzy communities or overlapping communities can be tested. 
\end{itemize}

Drawbacks:
\begin{itemize}
    \item Nothing guaranties that the networks and communities generated are realistic. This means that some algorithms might be more (or less) efficient on these generated networks than on real ones.
    \item This category of evaluation aim at finding a universally most efficient algorithm. By varying the parameters of the network, it might be possible to refine this classification, some algorithms being more efficient for dense networks, or small communities for instance, but the algorithms efficiencies are not evaluated on the particular graph that one wants to study.
\end{itemize}

\subsection{Evaluation on real networks with ground truth}
One solution to the problem of unrealistic networks and community structures of generated networks is to work with real networks and real communities. This was the idea of the first evaluations, using small networks such as the Zachary karate club \cite{zachary1977information} or Lusseau’s dolphins’ network \cite{lusseau2003emergent}, on which the communities found can be compared to a known real decomposition, or be studied graphically or intuitively. However, transposing this method to networks with a larger scale wasn't possible for a long period. Recently, some adequate networks were proposed, such as in \cite{yang2012defining,leskovec2010empirical}. In \cite{cazabet2012automated}, a slightly different approach is proposed: instead of comparing partitions to \textit{a priori} ground truth, experts assign relative and absolute scores to several solutions on a same network.

However, these methods also have weaknesses, as discussed in \cite{hric2014community} for instance. The problem is that this approach compares solutions that are purely topological with ground truths that depend on much more factors. It is for instance possible to have a ground truth community composed of several connected components, a situation that does not make sense on a topological perspective.

Other advantages and drawbacks are similar to those of the method using generated networks. To put it in a nutshell, whereas the network properties are no longer a concern, the solution of reference becomes less reliable, and this approach can only be used to pick an universally best performing method.

\begin{figure*}
        \centering
        \begin{subfigure}[b]{0.5\textwidth}
                \includegraphics[width=\textwidth]{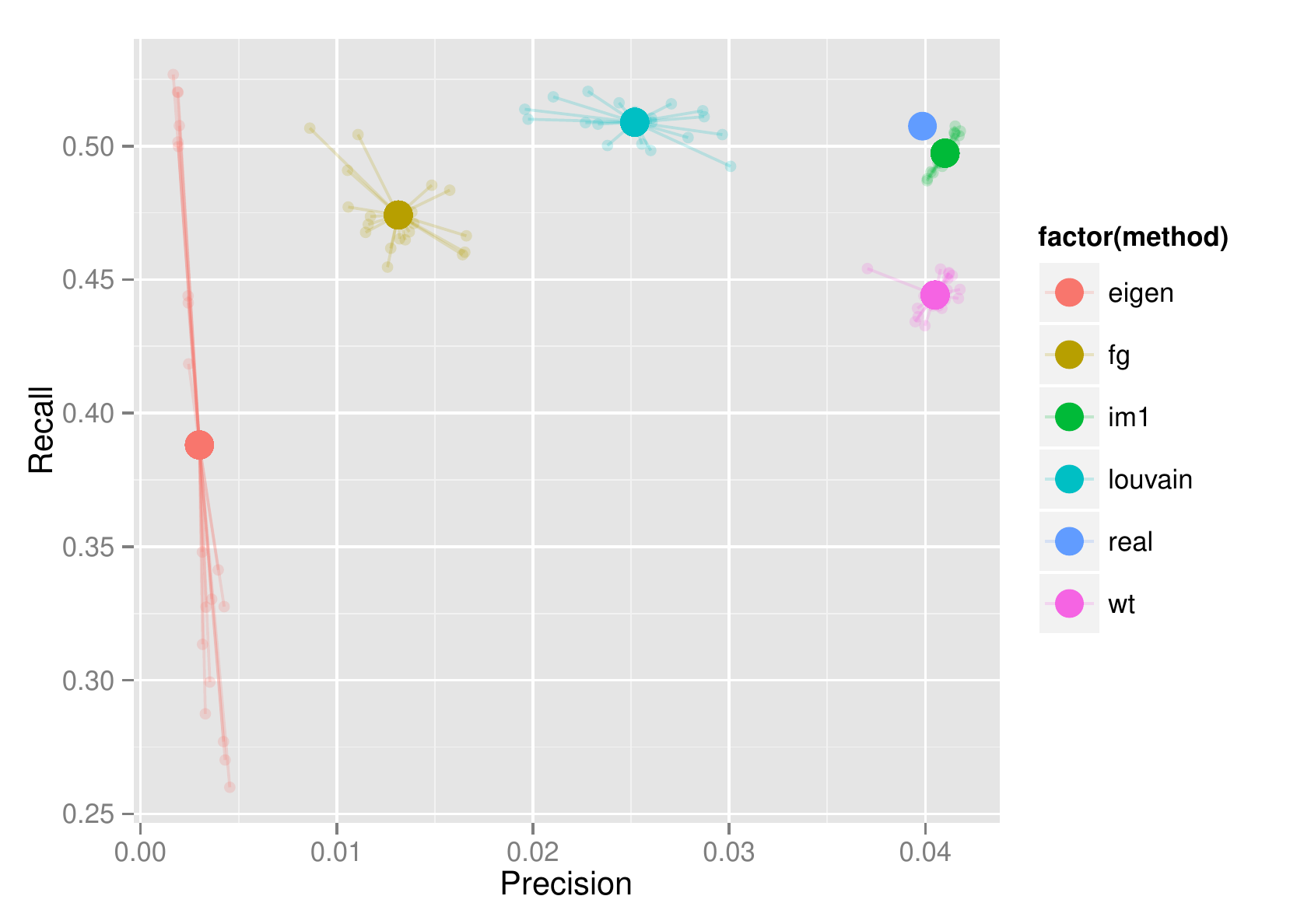}
                \caption{LFR, $\mu$=0.5 }
        \end{subfigure}%
        ~
        \begin{subfigure}[b]{0.5\textwidth}
                \includegraphics[width=\textwidth]{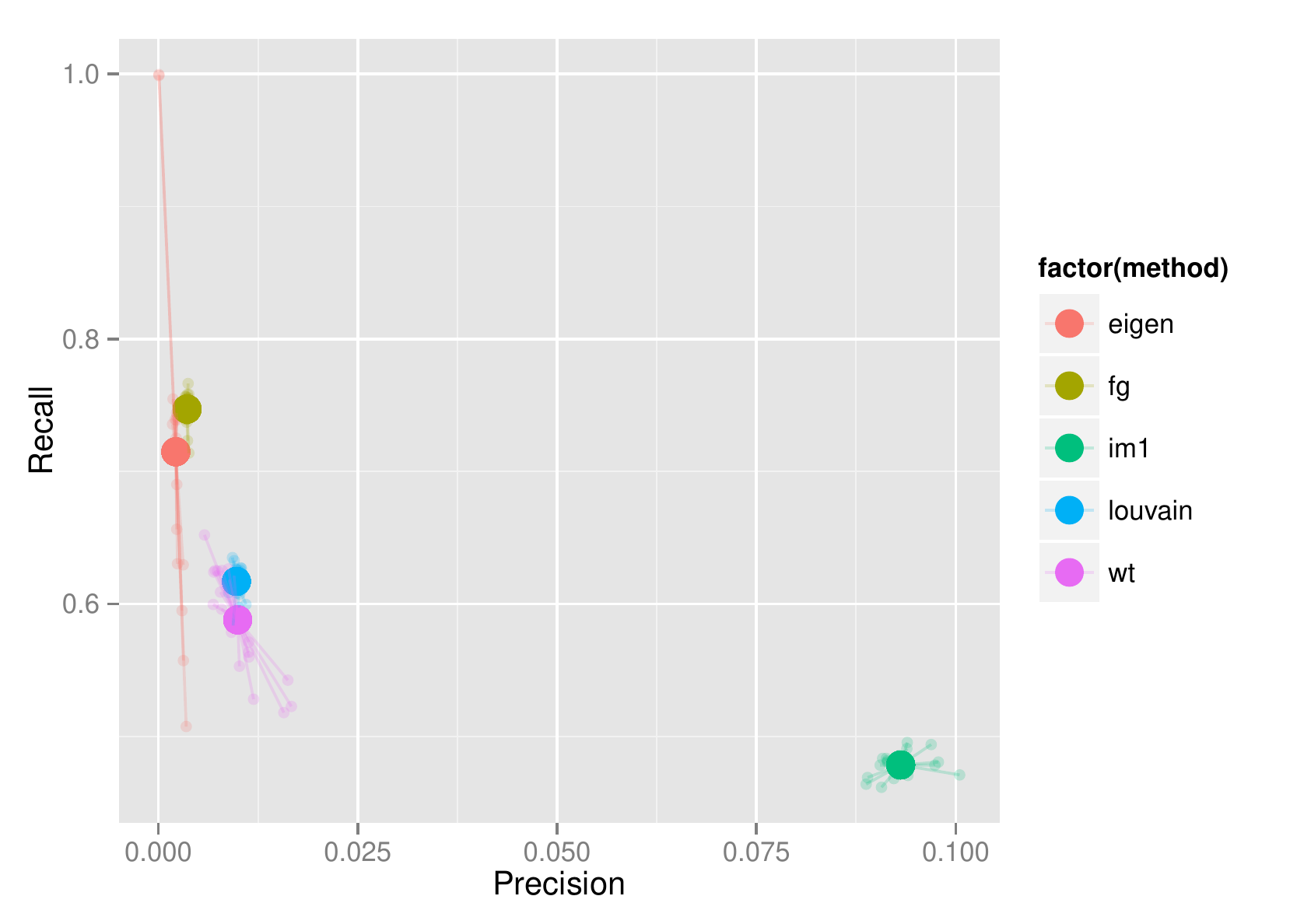}
                \caption{DicoSyn}

        \end{subfigure}%
    \caption{Visualization of Precision and Recall for several algorithms}
        \label{fig:PrecisionRecall}

\end{figure*}

\subsection{Potential advantages of a multiple-criteria quality function}
As we have seen in the previous section, using a quality function to evaluate community structures has the important advantage of evaluating partitions on a given network of interest, instead of searching for an universally superior algorithm. In the most common application of community detection, one wants to study a given network, knows several methods applicable to it, but does not know which one to use. The quality function can tell him which algorithm is the most efficient on his network. However, we know that with a single metric, the choice will be arbitrary, because it is possible on networks to have several relevant community structures of different granularities, and one of these solutions will, in the general case, have a higher quality score than the others. By using a multiple-criteria function and the notion of Pareto optimality, we will be able to identify several potentially relevant solutions, defined as all solutions present on the Pareto frontier. The property of these solutions is that no other solution is superior to them in all considered criteria. The preference for one of these solutions relatively to others can still be decided by attributing weights to the considered criteria. In the following sections, we will propose some possible relevant criteria.

\section{A first approach: Community Detection as a classification problem}
To propose a new method for community evaluation, we need to go back to the definition of what is a good community. The consensual definition, widely accepted, is that a good community structure must identify groups of nodes that are well connected between themselves, while having few edges between the groups.

The key idea that we will explore in this paper is that communities are defined as a trade-off between two objectives, having communities as dense as possible, and as well separated as possible from the rest of the network. 

Creating well-separated communities, without other constraint, is simple. By increasing the size of communities, we can add as many edges as we want in them. The optimal solution for the separation objective is to define communities as the connected components of the network, resulting in the absence of edges between communities.

In a similar fashion, without other constraint, it is straightforward to define communities as dense as we want. The optimal solution for this objective, reachable in any network, consists in communities composed of cliques, of size 2 in the worst case, plus a certain number of singletons. All non-singleton communities have a density of 1.

Between these two extreme but uninformative solutions lie the sought ones. 


This idea was already present in the graph-partitioning problem, which consist in optimizing the separation while fixing the size of communities. As the intern density depends on the number of intern edges and the size of the communities, this has the consequence of fixing a lower bound on the value of the intern density.

This idea is again central in the common definitions of communities, such has the conductance \cite{leskovec2009community}, defined as the ratio of extern edges over intern edges, or the modularity, in a more indirect fashion.

However, when comparing partitions, these two opposite objectives are usually merged in a single metric, to determine which solution is better than all the others.
As discussed earlier, this is often not pertinent in the case of community detection, as several meaningful levels of decomposition might exist.

The problem is to decide which metrics can we use to meaningfully represent the separation and the definition. In the coming chapter, we will propose a first simple approach, and discuss its strengths and weaknesses.

\subsection{Precision and Recall}
One of the most common uses of a twofold metric to evaluate the result of an algorithm can be found in the classification problem, through the usage of Precision and Recall. The first approach we propose is simply to consider the problem of community detection as a classification problem, allowing us to evaluate it as such.

We can define the problem as follows: for an undirected, not oriented graph $G =(V,E)$, the instances to classify are all the \textit{vertex pairs} of the network, $VP = \{(i,j):i\in V, j \in V, i \neq j\}$.  

The two categories are: \\
EDGE : \{$vp \in VP : vp \in E$\} \\
NOTEDGE : \{$vp \in VP : vp \notin E$\} \\
A community detection algorithm is therefore seen as a classifier to recognize vertex pairs that are the more likely to belong to $E$. This definition makes sense relatively to the definition of community detection: communities must be dense, and few edges must fall between communities, a community detection tries to leave as less edges outside of the communities, while having as less non-edges inside communities as possible.

To keep close with the classification problem, we split $E$ in a training set and a validation set. In our experiments, we create a graph $G_{training}$, corresponding to an original graph $G$ from which we remove a percentage $P$ of its edges.

These removed edges $RE$ constitute part of the validation set $VS$. However, the validation set must be representative of the training set used by the algorithm, which was composed of vertex pairs, not only of edges of $G$. Therefore, our validation set must be composed of both. The vertex pairs not belonging to $E$ are essential to compute the rate of false positives, edges that fall inside communities but do not belong to $G$. The total number of vertex pairs in the validation set is equal to 
$|VS| = P*|VP|$, and is composed of $\{RE \cup FPT\}$, with $FPT$ a random selection of potential edges taken from $\{VP \setminus E\}$, with size $|FPT| = |VS|-|RE|$

Precision and Recall are computed according to their usual definition. As a reminder, they are defined as: \\
$Precision = \frac{TP}{TP+FP}$ \\
$Recall = \frac{TP}{TP+FN}$\\
with TP = True Positive, FP = False Positive and FN = False Negative

\begin{table}[htbp]
\begin{center}
\begin{tabular}{rcl }
\hline
$C$ & : & set of all communities\\
$n$ & : & number of nodes in a graph\\
$m$ & : & number of edges in a graph\\
$l_c$ & : & number of edges inside community $c$ \\
$l$ & : & number of edges inside all communities\\
$l'_c$ & : & number of edges inside community $c$ according \\
 &  &  to a null model \\
$l'$ & : & number of edges inside all communities according \\
 &  &  to a null model \\

$k_c$ & : & number of vertex pairs inside community $c$ \\
$k$ & : & number of vertex pairs inside all communities \\
$p$ & : & number of vertex pairs in the whole graph\\

\hline
\end{tabular}
\end{center}
\caption{Notations}

\label{tab:definition}
\end{table}

\subsubsection{Network interpretation of Precision and Recall}
In the context of community detection, Precision and Recall can be written as a more traditional, network-centered approach. To simplify, we will not make a difference between the test set and the whole network. On the whole network, the Recall corresponds to the fraction of all edges that fall inside communities. This value is often used in evaluating community detection, for instance in the Modularity, defined as the difference between this observed ratio and the expected ratio in a null model.

Precision corresponds to the ratio of edges inside communities over the number of vertex pairs inside communities, i.e the global density of communities. More formally, on the whole network:
\[
Precision = \frac{l}{k} ,  Recall = \frac{l}{m}
\]

To keep the equations simple, we will use the notations of Table \ref{tab:definition} for all equations relative to communities.

\subsubsection{Precison and Recall to evaluate communities}

Fig. \ref{fig:PrecisionRecall} is an illustration of the results with five widely used algorithms, on a real graph and a generated graph produced by using the LFR benchmark \cite{lancichinetti2008benchmark}. 
The networks used are:
\begin{itemize}
\item LFR, $\mu=0.5$: a network generated with the LFR benchmark, with standard parameters and a mixing parameter $\mu = 0.5$. Communities are still well defined, but most algorithms already fail to identify them. $n=5000$, $m=50000$

\item DicoSyn \footnote{Network available at the following address: \url{http://dx.doi.org/10.5281/zenodo.12453"}}\cite{gaume2008toward}: a synonymy network, nodes represent verbs and edges proximity in meaning. We chose this network because it is easily interpretable, and one can observe by comparing different algorithms that partitions with high modularity seem less relevant than some solutions of lower modularity. $n=9146$, $m=51423$
\end{itemize}

The methods used are:
\begin{itemize}
\item \textbf{eigen}: The eigenvector decomposition method described in \cite{newman2006finding}, as implemented in the igraph package \cite{Csardi:2006uq}.
\item \textbf{fg}: The fastgreedy modularity optimization method, described in \cite{clauset2004finding}, as implemented in the igraph package. It provides a hierarchical decomposition.
\item \textbf{wt}: The walktrap algorithm, described in \cite{pons2005computing}, as implemented in the igraph package.
\item \textbf{im}: The infohiermap method, described in \cite{rosvall2011multilevel}, as implemented by the authors. Can identify several levels.
\item \textbf{louvain}: The louvain method, described in \cite{blondel2008fast}, as implemented by the authors. Can identify several levels.
\end{itemize}

For each network, we generated 20 different test sets, and run each algorithm on each of these test sets. Individual results are displayed as a small dot, and a large dot corresponds to the average values. For methods producing several solutions, we consider only the default one.

On the generated graph, we can observe that only two methods, infomap and louvain, are on the Pareto frontier, the other ones being outperformed on both aspects at least by infomap. The solution proposed by the louvain method has a higher Recall than the infomap method, but a lower precision. The correct decomposition is also on the Pareto frontier.

On the real graph DicoSyn, it seems more difficult for a solution to outperform most others. Instead, each method is superior to other ones on one aspect, but inferior in another one. The eigen method is the only one to be outperformed. 

These examples illustrate the interest of a multi-objective approach:
\begin{itemize}
    \item It is possible to compare algorithms on a particular graph, and not only on test graphs
    \item It is possible to eliminate some methods as Pareto dominated
    \item It is also possible to keep several partitions as potential solutions, and to choose the most interesting one depending on our objective
\end{itemize}

\subsubsection{Limits of the Precision and Recall method}
By using the typical definition of Precision and Recall, we have a first solution for the evaluation of the quality of partitions. However, this method has clearly some weaknesses. First, community detection is not usually defined as a classification problem, and there is no guaranty for these metrics to correctly recognize good communities.

The second weakness is that we can easily add a large number of partitions on the Pareto frontier, even though these partitions are clearly not relevant in term of communities. We can show that, starting from a Pareto optimal partition $P$, we can in most cases generate a new valid partition $P'$ with either a higher Precision or a higher Recall. A higher Recall can be obtained by merging communities having edges between their nodes (increase in $l$, $m$ stays constant). A higher Precision can be reached, for instance, by splitting sparse communities in cliques of size at least 2. Note that the resulting solutions are not necessarily Pareto optimum for the graph, it could be possible to find a solution Pareto dominant to them, but they are not Pareto dominated by $P$.

To put it in a nutshell, the problem of this method is that it is too simple to generate irrelevant solutions on the Pareto frontier. In fact, this problem is linked to the size of the communities found: a partition composed of large communities will tend to have a high Recall and a low Precision, and \textit{vice versa}. In the next section, we will propose a solution taking into account this problem of the size of communities.

\begin{figure*}
        \centering
        \begin{subfigure}[b]{0.5\textwidth}
                \includegraphics[width=\textwidth]{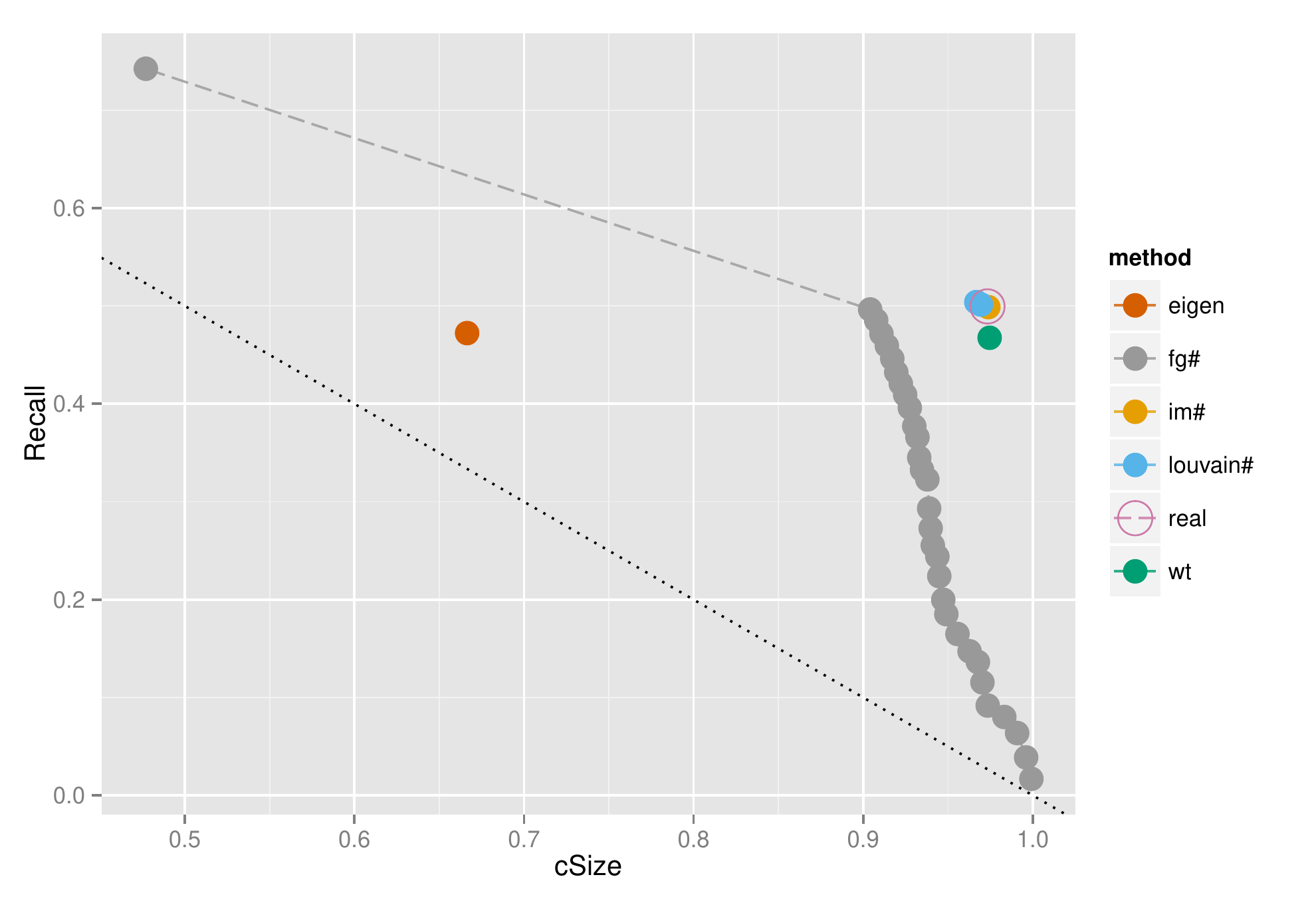}
                \caption{LFR, $\mu=0.5$ }
        \end{subfigure}%
        ~
        \begin{subfigure}[b]{0.5\textwidth}
                \includegraphics[width=\textwidth]{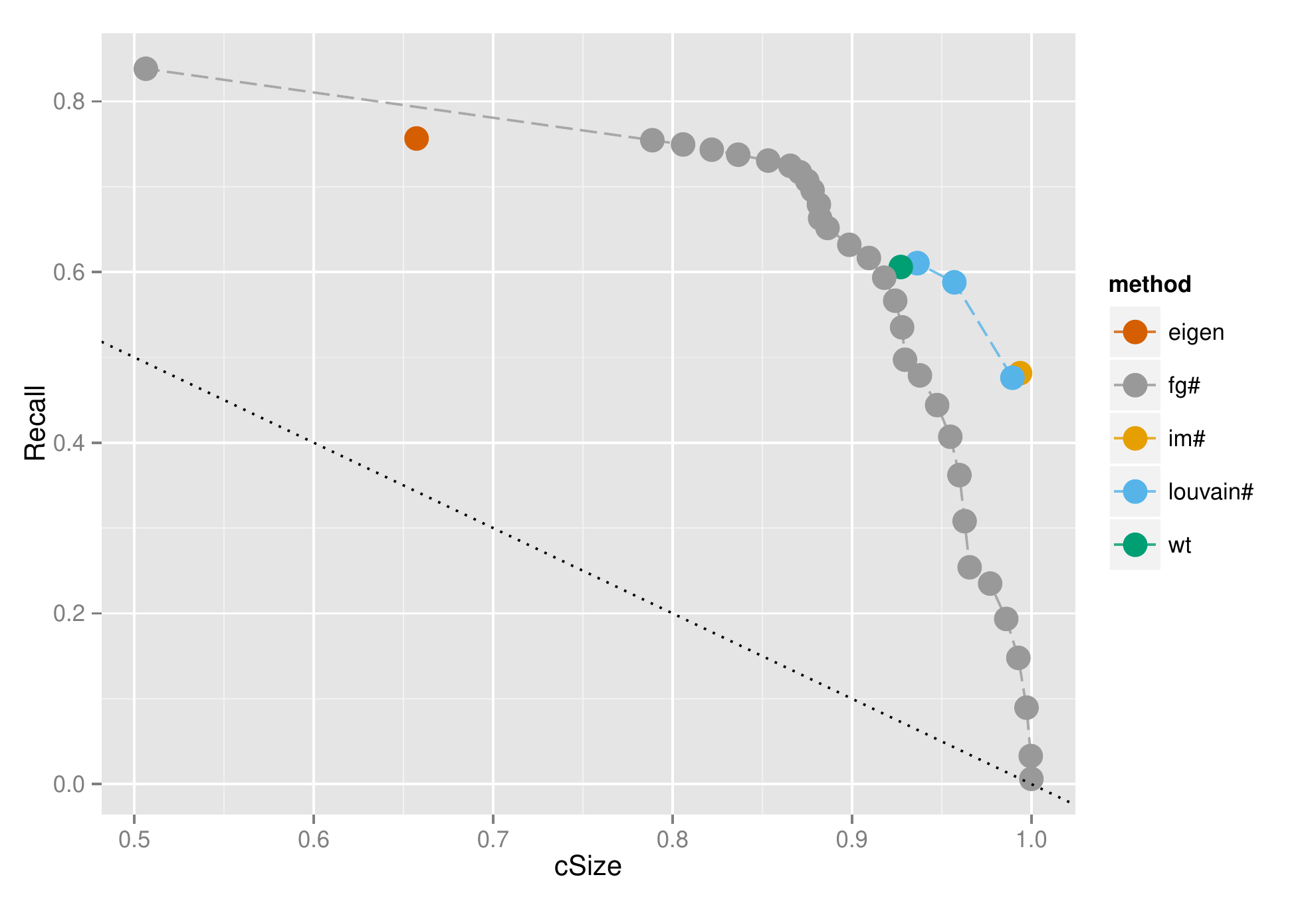}
                \caption{DicoSyn}

        \end{subfigure}%
    \caption{Visualization of the ratio of intern edges and size of communities}
        \label{fig:sizeSeparationNoMod}

\end{figure*}

\subsubsection{Single run versus several runs}
In this first approach, because we wanted to stay as close as possible from the classification problem, we used a training set and a test set. However, as our problem is not really a classification problem, there is no need to do so, and we could compute the values of Precision and Recall on the whole network. For the other solutions presented on this paper, we will directly compute solutions on the whole network, and compute our criteria accordingly. However, we can note that the idea of evaluating several runs on a slightly randomized version of the network can nevertheless be interesting, because of the known problem of the instability of the algorithms\cite{aynaud2010static}: a same algorithm, when confronted to two slightly different versions of a same network, can converge to very different partitions.

\section{Size, Separation and Modularity}
The method presented in the previous section has the drawback of not being directly grounded on the traditional works on community detection. In this section, we will propose another approach which is based on a generalization of the graph partitioning problem, and we will show that this approach is directly related to the most widely used quality function in community detection, Modularity.

\subsection{Graph partitioning generalization}
The ancestor of community detection is the problem of graph partitioning. This problem can be expressed in the following manner: for a given network and a given number of clusters of similar sizes, the best partition is the one that will result in a minimum number of edges laying between clusters. It is necessary to fix the number and size of the objective clusters, otherwise minimizing inter-cluster edges is achieved by a trivial solution, such as leaving only the node of lower degree in a community and all other nodes in another. As the number of extern edges is the opposite of the number of intern edges, a typical measure of the quality of a partition for communities of fixed properties can be unambiguously defined as:
\[
Q_{partitioning} = 1 - \frac{m-l}{m} = \frac{l}{m} = Recall
\]
Which is identical to the Recall defined in the first chapter. However, two partitions can only be compared according to this metric if they are composed of clusters of similar sizes. The problem of community detection can be seen as a generalization of graph partitioning, searching for the best partition not only for fixed properties of communities, but for the best solution considering all possible combinations of number and size of communities.

The two opposite metrics that we propose to use are therefore $Q_{partitioning}$ and an indicator of the size of communities, which corresponds to the difficulty of having intra-community edges. The more vertex pairs are inside communities, the easier to have edges inside communities. To represent the size of the communities, we want to use a metric between 0 and 1, with 0 corresponding to the largest communities, for which a maximal value of $Q_{partitioning}$ can be reached. We use the fraction of vertex pairs that fall outside communities:
\[
cSize = 1 - \frac{k}{p} 
\]

We can remark that this metric is related to diversity indexes: small values will correspond to larger, more uneven clusters than large values. Said differently, the closer we get to a single cluster containing all nodes, the less useful information the partition contains about the modular structure of the network.

Fig. \ref{fig:sizeSeparationNoMod} is a visualization of these two metrics on the same networks used in the previous section. This time, we display all levels of decomposition provided by hierarchical algorithms. One interesting property is that, if the graph was a random one, the proportion of edges inside communities will be linearly proportional to the proportion of vertex pairs inside communities. Therefore, on a random graph, there is a relation $Q_{partitioning} = 1-cSize$. We represent this relation as a straight dashed line in our graph.

Using this random case as a baseline allows us to balance the improvement in $Q_{partitioning}$ when augmenting the size of communities relatively to the mechanical improvement due to the higher number of vertex pairs between communities. By taking the difference between the separation produced by the partition and the baseline, we can have a measure of the improvement yielded by this partition. This function $Q'$ can be defined as:
\[
Q' = Q_{partitioning} - \frac{l'}{m} 
\]
This is a partial solution to the problem we encountered using the Precision/Recall approach: if we consider $Q'$, instead of the raw $Q_{partitioning}$ value, it is no longer possible to provide trivial solutions by merging communities of an initial partition, as these trivial solutions will come closer to the baseline, if they are not relevant, and therefore have a lower $Q'$.

\begin{figure*}
        \centering
        \begin{subfigure}[b]{0.5\textwidth}
                \includegraphics[width=\textwidth]{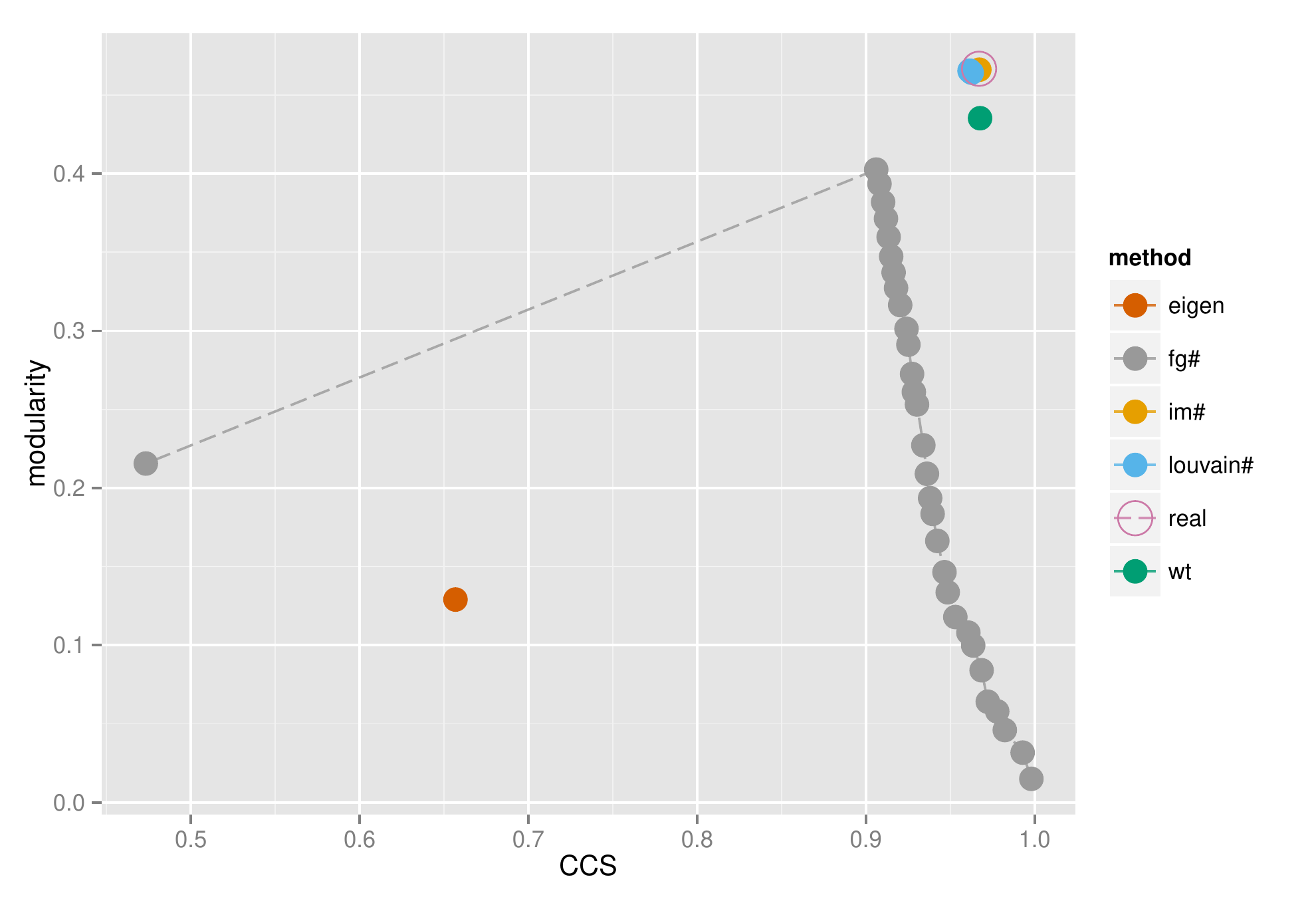}
                \caption{LFR, $\mu=0.5$ }
        \end{subfigure}%
        ~
        \begin{subfigure}[b]{0.5\textwidth}
                \includegraphics[width=\textwidth]{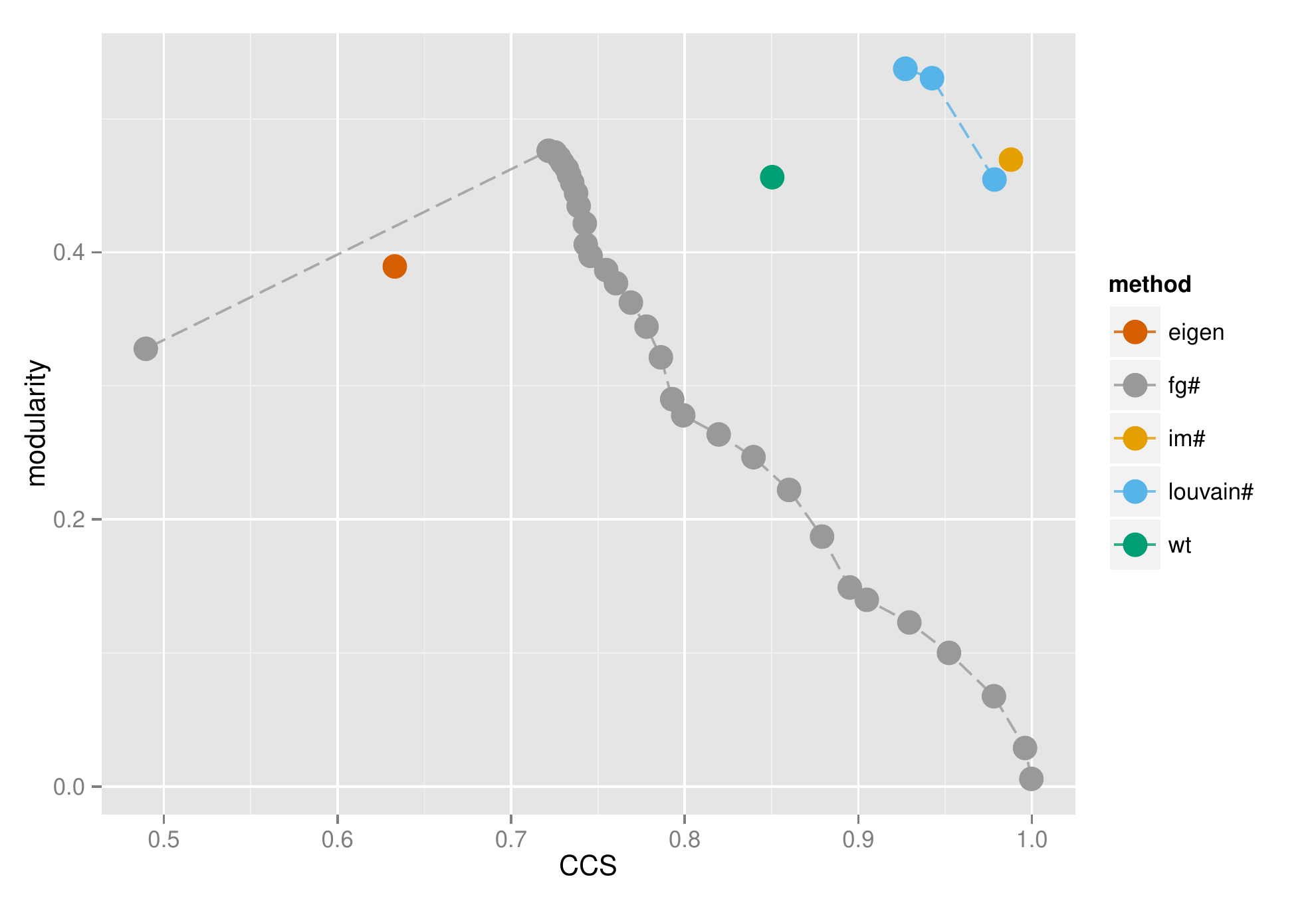}
                \caption{DicoSyn}

        \end{subfigure}%
    \caption{Visualization of the Modularity Decomposition Graph}
    
    \label{fig:modCCS}

\end{figure*}

\subsection{Relation with the Modularity}
Modularity is usually defined as a sum over all edges or a sum over communities. However, using our notation, it can also be written as: 

\[
Modularity = \frac{l}{m} - \frac{l'}{m}
\]
Where $l'$ corresponds to the expected number of edges according to a null model. As a consequence, if the same null model is used, $Q'$ is strictly equivalent to the Modularity. Previous works have shown that a better null model consists of a random network of same degree distribution than the original graph. We can adapt our solution to this improvement.


\begin{figure}[h]
        \centering
        \begin{subfigure}[b]{0.5\textwidth}
                \includegraphics[width=\textwidth]{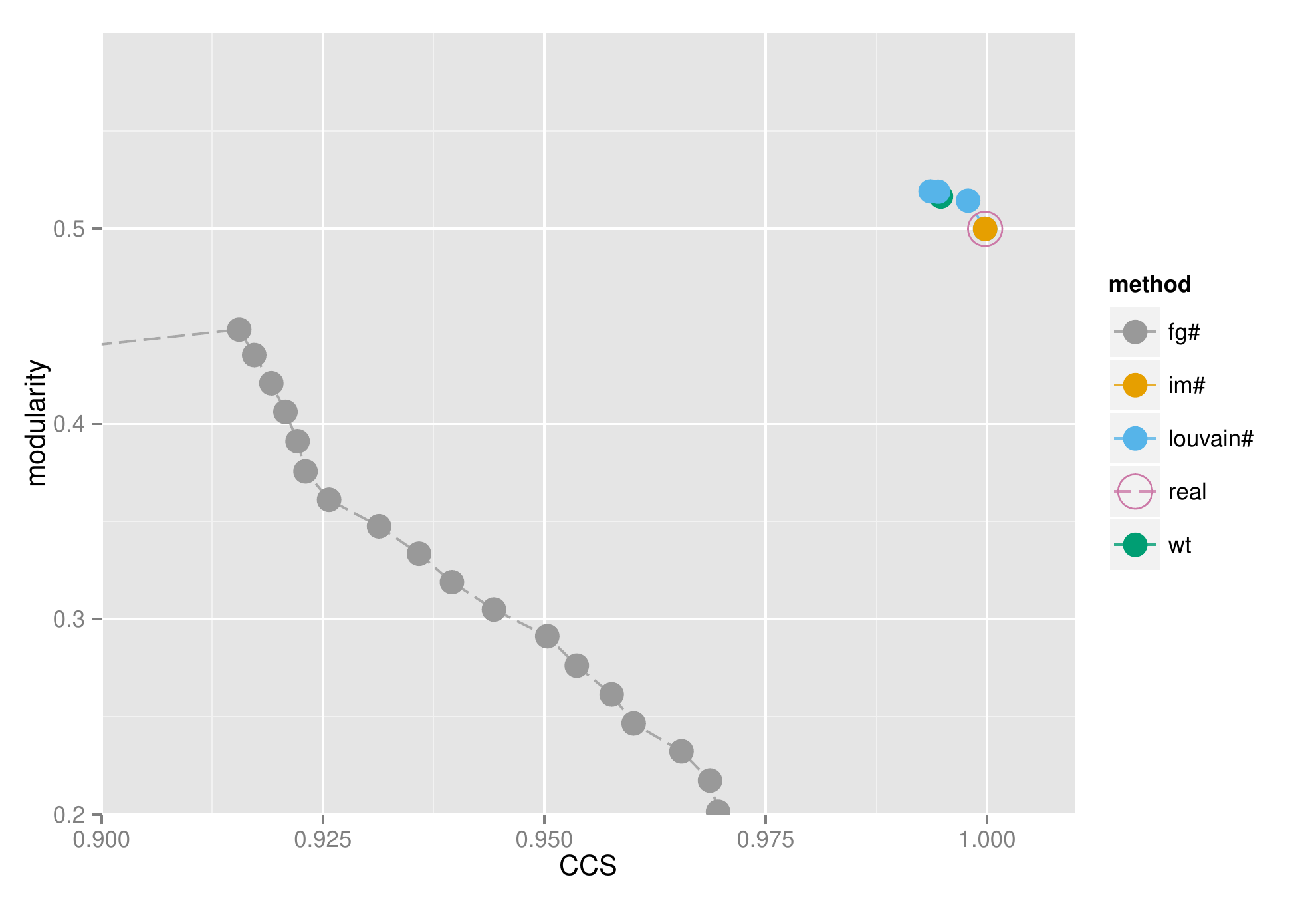}
        \end{subfigure}%

    \caption{Modularity Decomposition Graph on a generated network where the maximum of modularity do not match with the researched solution}
    \label{fig:killMod11}

\end{figure}

\subsubsection{Modularity Decomposition Graph}
Based on the previous observations, we can propose a variation of the $Q_{partitioning}$ and $cSize$ criteria. We already proposed to replace $Q_{partitioning}$ by $Q'$, and we have just seen that using the typical Modularity was an improvement over it. However, we must also change the $cSize$ accordingly. By using the property $\frac{k}{p} = \frac{l'}{m}$ in a random graph, we can define our modified $cSize$, that we call Corrected Community Size ($CCS$) as:
\[
CCS = 1 - \frac{l'}{m} = 1 - (Q_{partitioning} - Modularity)
\]
The idea of this measure is similar to the $cSize$, it represents how big communities are in term of the probability of containing links, but this time considering the degrees of their nodes. Fig. \ref{fig:modCCS} represents the score of partitions using the same algorithms and graphs than previously. A consequence of using these two metrics to define a Pareto frontier is that, if we know the solution of maximal modularity, it is not possible to find a value on the Pareto frontier with a value of $CCS$ below the value corresponding to the modularity optimum, that is to say, it is not possible to find a Pareto optimal solution with "larger" communities than the solution of maximal modularity. As a consequence, in the LFR example, most solutions become Pareto-dominated by the correct partition known by construction.

While working with networks generated by the LFR benchmark, we were surprised to observe that the correct solution was always the solution of highest modularity, despite the so-called resolution limit. This is because, with the parameters most commonly used in the literature, the communities are in the correct resolution for modularity. To avoid this bias, we generated a network with the LFR benchmark using the following parameters: 
\begin{itemize}
    \item Number of nodes: 50000
    \item $\mu$: 0.5
    \item Size of communities : [11,11]
    \item Average degree: 20
    \item Maximal degree: 20
\end{itemize}
Note that the LFR benchmark is not fully appropriate to generate this kind of large graphs with small, dense communities, because of its universal $\mu$ value for each node. This is the reason for our choice to generate cliques of fixed size, and to set the degrees of nodes accordingly. Although unrealistic, this is not a major concern for our purpose, as we are just interested in obtaining a clear community partition with suboptimal modularity score, and not to compare community detection algorithms on realistic networks.

The results of this graph are shown in Fig. \ref{fig:killMod11}.  We can observe that, although both Louvain and InfoMap found the correct decomposition, the Louvain method also identifies solutions of higher modularity, but of larger size. The fast greedy method identifies a solution with a Modularity score relatively close to the optimum, but very far in term of the size of communities.
\begin{figure*}
        \centering
        \begin{subfigure}[b]{0.5\textwidth}
                \includegraphics[width=\textwidth]{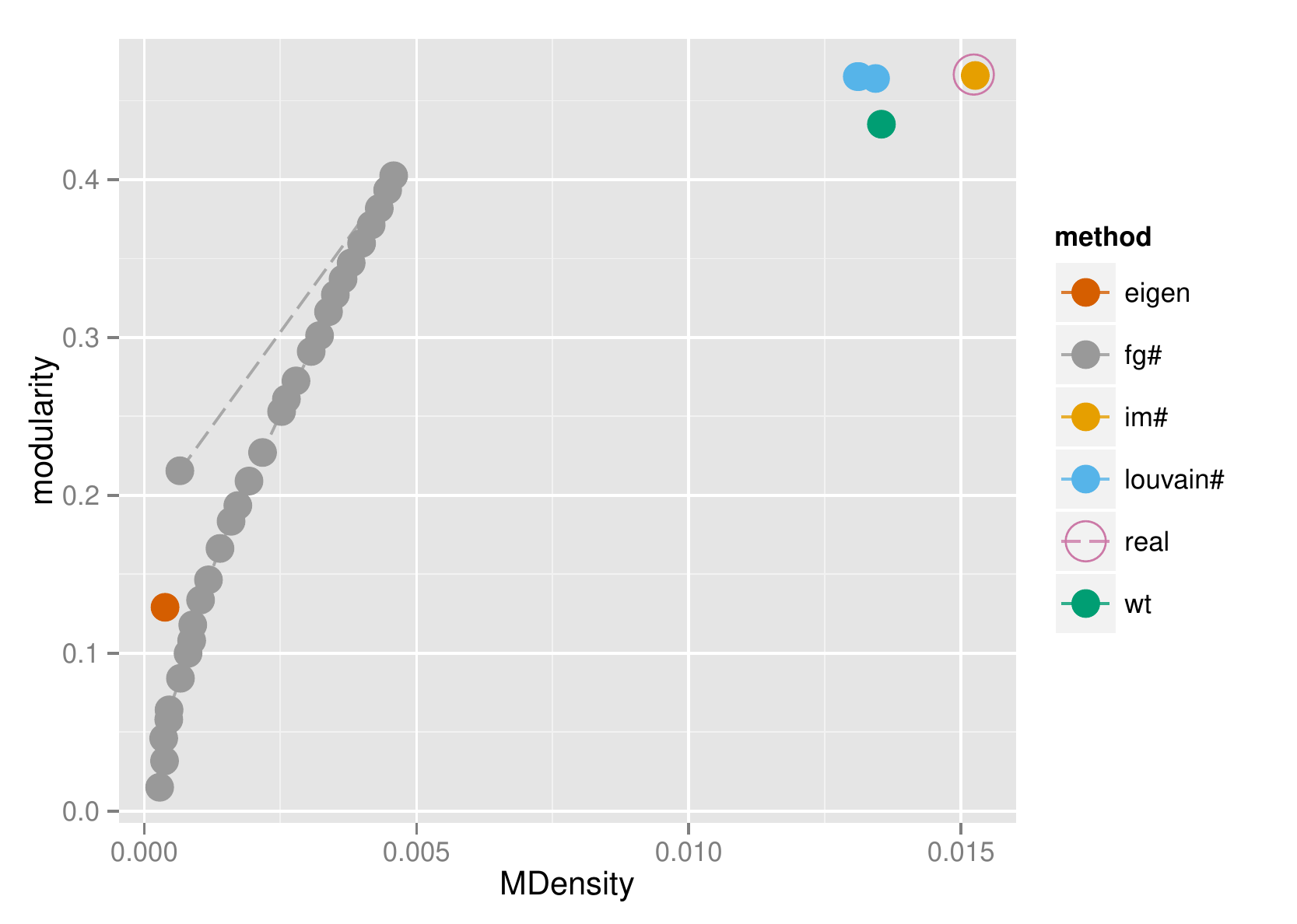}
                \caption{LFR, $\mu=0.5$ }
        \end{subfigure}%
        ~
        \begin{subfigure}[b]{0.5\textwidth}
                \includegraphics[width=\textwidth]{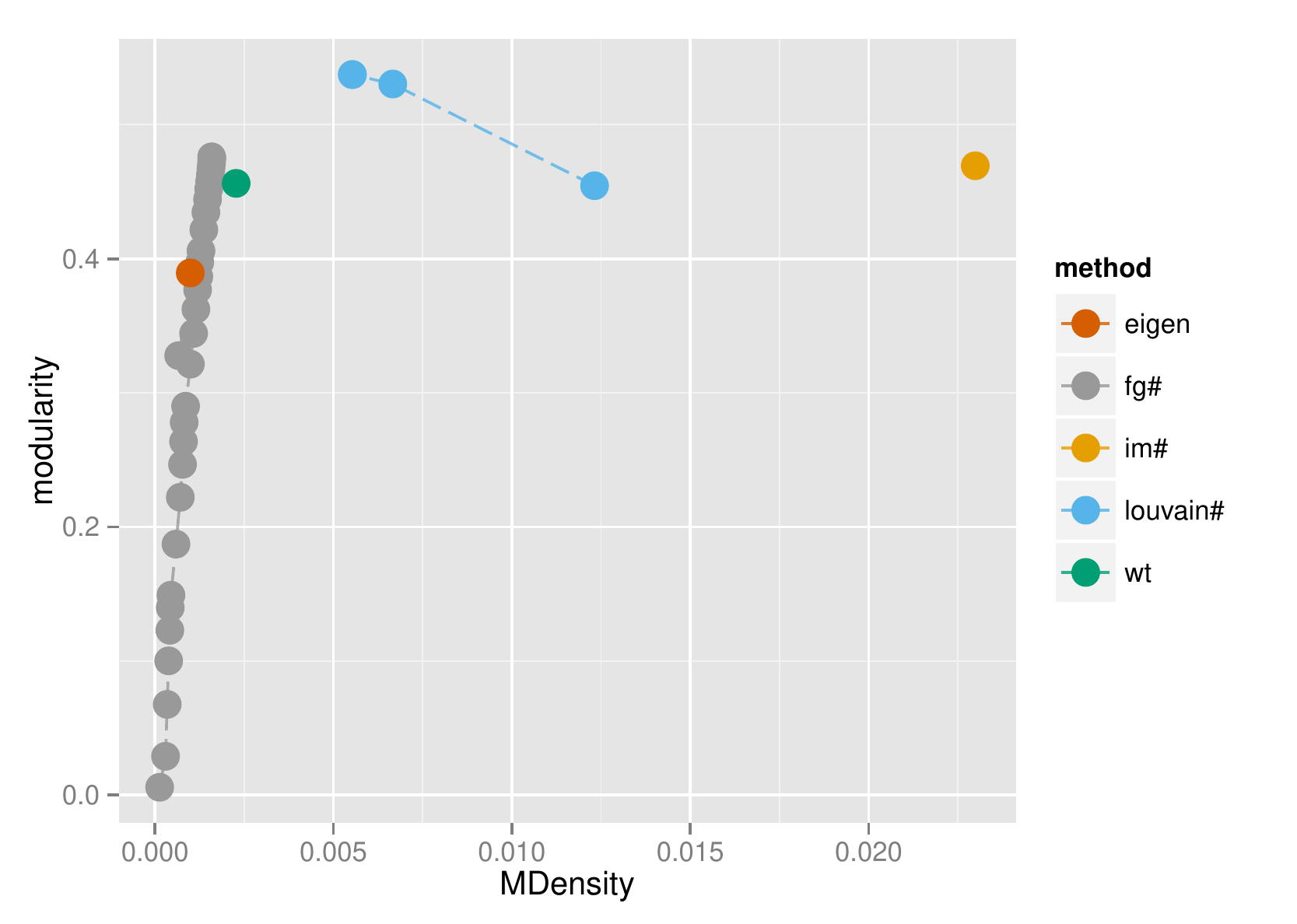}
                \caption{DicoSyn}

        \end{subfigure}%
        \label{fig:modCSS}
        
        \begin{subfigure}[b]{0.5\textwidth}
                \includegraphics[width=\textwidth]{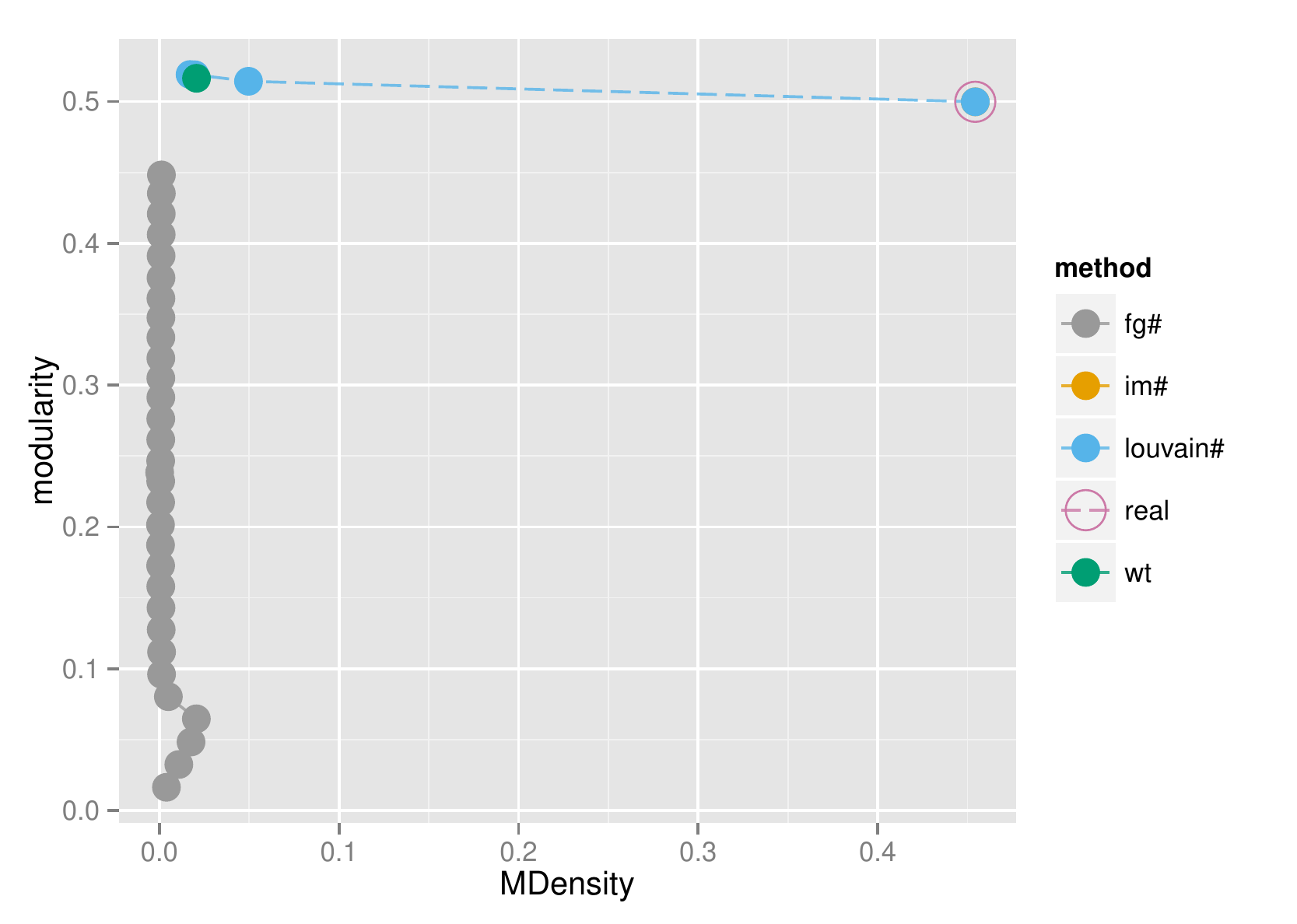}
                \caption{LFR, $\mu=0.5$, suboptimal Modularity}

        \end{subfigure}%
        \begin{subfigure}[b]{0.5\textwidth}
                \includegraphics[width=\textwidth]{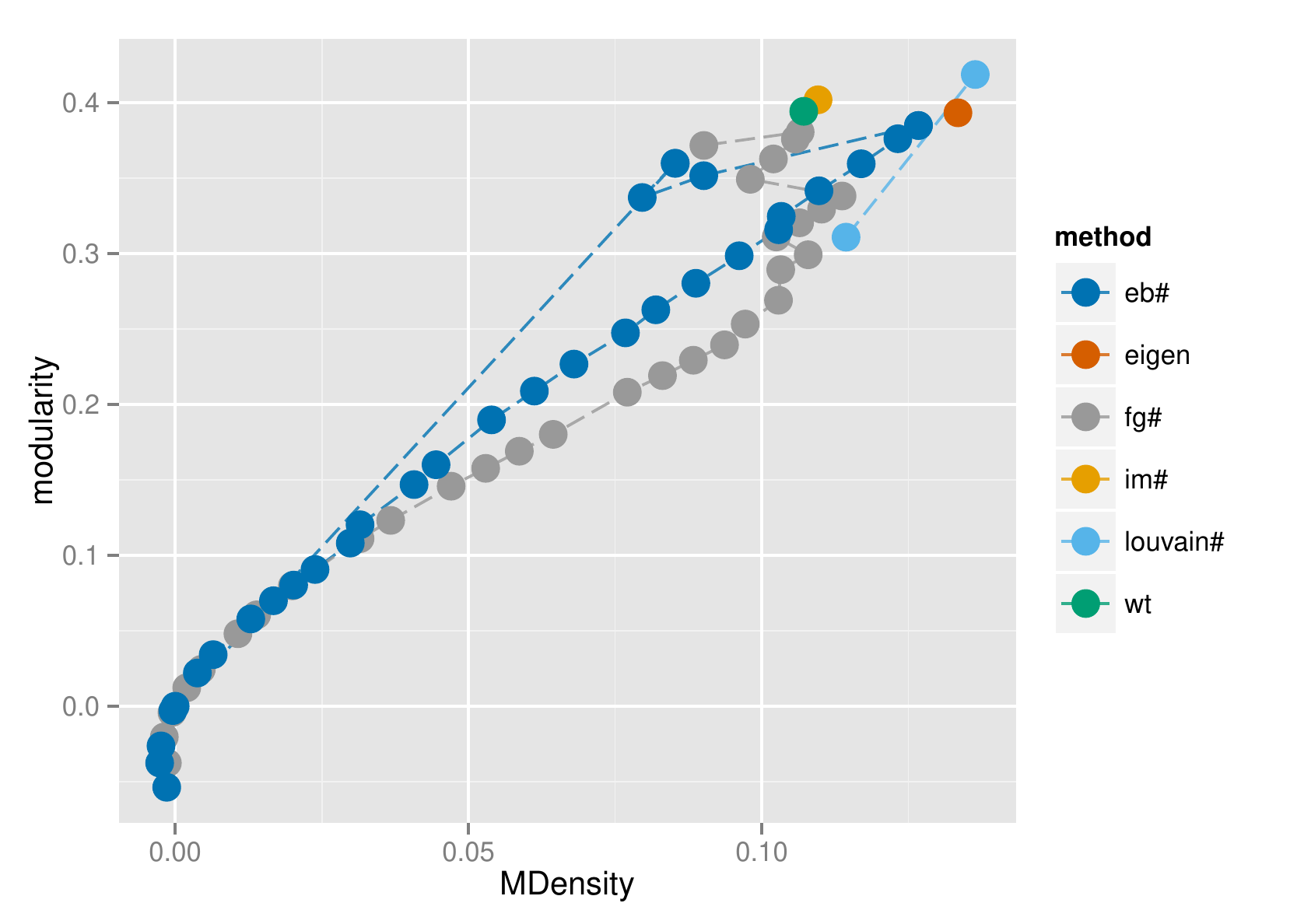}
                \caption{Zachary karate club}

        \end{subfigure}%
        
         \begin{subfigure}[b]{0.5\textwidth}
                \includegraphics[width=\textwidth]{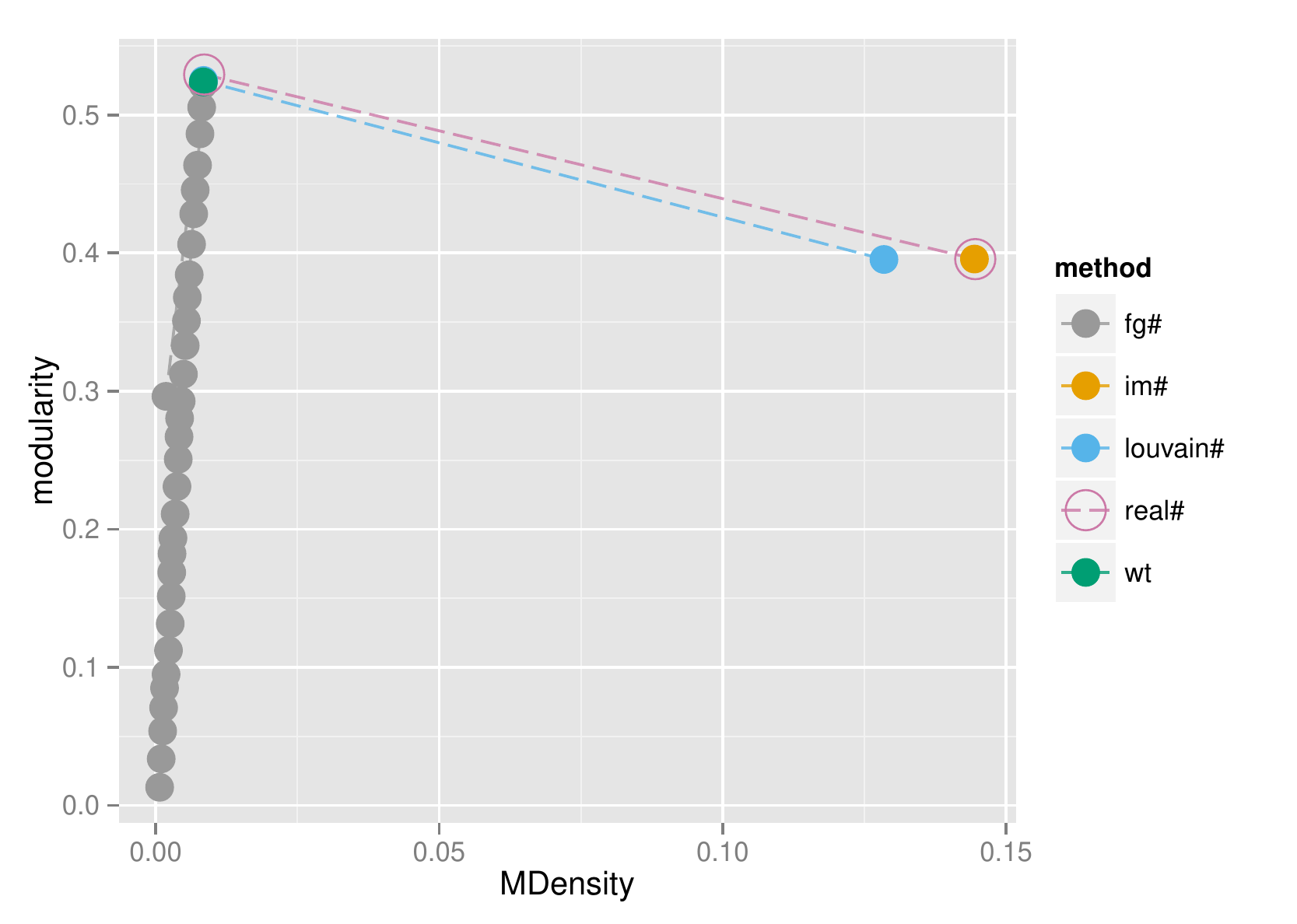}
                \caption{LFR $\mu=0.3$, 2 levels. infomap's solution is identical to the real decomposition of lower modularity, walktrap's solution is identical to the other real decomposition.}

        \end{subfigure}%

    \caption{Comparison of partitions using Modularity and MDensity}
        \label{fig:finalCompare}

\end{figure*}

\subsubsection{Limitations of the Modularity Decomposition Graph}
Compared to the first multiple-criteria approach that we proposed, the Modularity Decomposition has several advantages: it is directly based on the Modularity, a strongly established quality function for partitions, and it does not allow to add points on the Pareto front by generating arbitrarily large communities. Compared to using the sole Modularity, it allows differentiating between suboptimal but pertinent partitions --with a value of modularity below the optimum but corresponding to "smaller" communities-- and suboptimal partitions, yielding solutions not on the Pareto frontier.

However, this method still has some drawbacks: first, it is still possible to find trivial solutions by proposing arbitrarily small communities. Secondly, as seen on Fig. \ref{fig:killMod11}, this solution might not help us to prefer the correct solution over the solution of optimal modularity when relevant. Whereas the correct solution is on the Pareto frontier, the gain in CCS is of an order of magnitude comparable to the loss in the Modularity.

\section{Modularity and MDensity}
The Modularity is a method based on the comparison of the ratio of extern (or intern) edges to a null model. As so, this metric is clearly a descendant of the partitioning problem. However, compared to the traditional, informal definition of communities, which states a trade-off between clearly separated and well defined communities, it seems that the modularity alone is biased --for sparse networks-- toward an optimization of the separation of communities, at the cost of a poor definition. This phenomenon is a consequence of the limit of resolution of the modularity, and should be balanced by a symmetrical measure ensuring the optimization of the density of the communities. 

The modularity is a measure of the improvement of the separation of communities --the ratio of inter-community edges-- over a null model. We introduce the Module-Density, or MDensity, which is the improvement of the density of communities over a null model, pondered by the separation.

\subsection{Introducing MDensity}

We can start by computing the overall improvement in modularity over a null model, defined as the sum for each community of its gain in density pondered by its relative size in terms of number of pairs:
\[
DensityGain = \sum _{ c=1 }^{ |C| }{ \left( \frac { l_{ c } }{ k_{ c } } -\frac { l'_{ c } }{ k_{ c } }  \right) * \frac { k_{ c } }{ k } }  = \frac { l -l' }{ k }
\]

However this function has trivial maximal solutions, communities composed of cliques, we therefore balance it by the ratio of intern edges inside communities,  :
\[
MDensity = \frac { l -l'}{ k } * \frac { l }{ m } 
\]

We can then conveniently express the MDensity from the Modularity, which gives us another interpretation of the MDensity as the modularity balanced by the density.
\[
MDensity = \frac{l}{k}* Modularity
\] 

However, one of the interesting features of the Modularity is that it uses a null model based on the preservation of the degree distribution. We want to also integrate this feature in the MDensity. The intuition is that a higher density can be reached more easily between nodes of high degrees, and is therefore less significative than between nodes of low degrees. We have defined $l'$ as the number of edges expected in communities. As edges are distributed at random, we can define $k' =\frac{l'*p}{m}$, the degree-corrected number of vertex pairs inside communities based on the chosen null model. Our final definition of MDensity is now:
\[
MDensity = \frac { l -l'}{ k' } * \frac { l }{ m } 
\] 

To understand the difference in nature between MDensity and Modularity, we can compare what does a maximal score corresponds to for these two metrics. In both cases, a maximal score can be obtained only if the difference between the number of intern edges and the expected number of edges according to the null model is maximal. If we consider an infinite network with communities of finite size, the number of expected edges will tend to zero, maximizing this difference. As the sum of this difference is divided by the total number of edges for the modularity, any decomposition of an infinite network in connected components of finite size will result in a maximal score of Modularity of 1, whatever the properties of these connected components. We can imagine them as arbitrarily long chains with arbitrarily large cliques at their extremities for instance, which is probably not what most people will recognize as good communities. On the contrary, for the MDensity to be equal to 1, it is necessary that all communities have a density of one. As a consequence, a perfect score of MDensity can be reached in an infinite graph only if we can find communities defined as cliques without any links between them.

Fig. \ref{fig:finalCompare} presents the results on the same graphs as previously, plus the Zachary karate club and a generated network with 2 hierarchical levels. We can see that using these metrics often allows us to eliminate much of the proposed partitions, as not Pareto optimal. Even though the fast greedy Modularity optimization method proposes a complete hierarchy of solutions, none of them manage to be on the Pareto frontier. Each network result seems relevant:
\begin{itemize}

\item For the LFR benchmark, the correct solution is, as expected, on the top right area, only challenged by the infomap solution. 
\item For the DicoSyn network, the infomap solution, qualitatively identified as more relevant, sees its slightly lower Modularity compensated by a large MDensity score. The profile of this plot is interestingly similar to the next one.
\item For LFR with suboptimal Modularity solution, the researched solution is the only one to have both a high Modularity and MDensity. 
\item  For the Zachary karate club, as the network is small, we also represented the partitions obtained by the original edge betweenness algorithm by Girvan and Newman \cite{girvan2002community}, identified as $eb$. The solution of maximal modularity found by the louvain algorithm corresponds to the partition in 4 communities, often considered the most relevant. Despite the large quantity of partitions considered in this case, this solution is the only Pareto optimal one.
\item For the generated network with hierarchical communities, the two solutions are on the Pareto frontier, one with higher modularity and the other with higher MDensity. Different algorithms find one or the other of these solutions, illustrating the interest of our 2-criteria approach.
\end{itemize}

\subsection{Relation to Precision/Recall}
If we step back to our first approach, using Precision and Recall, and we compare to the couple Modularity and MDensity, we can now observe that there is a relation between them. The Modularity, as already stated, corresponds to the improvement between the observed value of Recall and the Recall of a similar partition in a random network. The MDensity can be defined as the Precision multiplied by the Modularity.

\section{Identifying best solutions}
 Contrary to the previous approach, in the generated graphs with a solution of suboptimal modularity, the gain of the correct solution in MDensity is much higher than the gain in Modularity of the other solutions of the Pareto frontier. We can use this fact to propose a combination of the two criteria in a single quality function. Of course, by doing so, we fall again in the problems of single criterion metrics described in introduction. As a consequence, we prefer to include in this metric a parameter $\alpha \in [0,1]$, which describes the relative importance we attribute to each criterion. Our combined quality function, 2FQ, for Two Fold Quality, is defined as:
\[
2FQ = \alpha Modularity * (1-\alpha) MDensity
\]
On the network of suboptimal modularity, we can now compute that the researched partition has the highest value of 2FQ for $\alpha \in [0,0.965]$, which means that it will be considered as the best solution, unless we choose $\alpha > 0.965$, corresponding to a choice mostly considering the separation of the communities, and not their density. We have to stress that the intervals of $\alpha$ corresponding to maximal 2FQ are only relative to the tested partitions, and that other partitions might exist that would completely modify them. The length of the interval also has no meaning, unless we know all the solutions on the Pareto frontier.

\section{Conclusion}
In this article, we presented a new approach to evaluate and compare partitions in communities. This method is grounded on the usual definition of communities, defined as a trade off between the "internal definition" and the "separation" of the communities. It makes use of an already widely used metric, the Modularity, and we can also relate it to the Precision and Recall approach in classification. Because it has two criteria, it allows identifying several relevant partitions. However, as both criteria do not have trivial solutions, it can drastically limit the number of partitions considered as relevant.

This method opens several possibilities for future work, among them:
\begin{itemize}
\item Adapt it to overlapping communities. This is not trivial, but some works already exist to adapt modularity to overlapping communities, such as \cite{nicosia2009extending}. 
\item Adapt existing algorithms of Modularity optimization for 2FQ optimization, and test the efficiency of such a method on simulated and real benchmarks.
\item Explore in details the properties of the Pareto maximal partitions, in the case of clearly or weakly defined communities.
\end{itemize}
To conclude, we want to stress the importance of comparing several partitions. Community detection is applied in many fields and for many purposes, but one popular usage is to use it on an existing network on which one wants to gain insights, and to interpret the partition found as being an intrinsic property of the studied network. For instance, one can study the sizes of the communities --or the distribution of their sizes, the number of inter-community edges, or more generally interpret the meaning of such and such nodes being clustered together. This is a perfectly legitimate practice, however, for this usage, each particular community detection algorithm having its own definition of what is a community, it appears important to take into consideration several relevant solutions, and to check if the observations we get on one partition are confirmed by others. For instance, it appears clearly from our observations and from the resolution limit that methods based on modularity optimization tend to find large, sparse communities when applied to large networks, while other methods could find a completely different but relevant solution composed of much smaller communities. We hope our multiple-criteria approach can be used in such cases to consider several partitions, not by applying randomly some algorithms, but by picking the most relevant partitions among several ones.
\vspace*{-3pt}   


\bibliography{biblioAll}

\begin{thebibliography}{27}%
\makeatletter
\providecommand \@ifxundefined [1]{%
 \@ifx{#1\undefined}
}%
\providecommand \@ifnum [1]{%
 \ifnum #1\expandafter \@firstoftwo
 \else \expandafter \@secondoftwo
 \fi
}%
\providecommand \@ifx [1]{%
 \ifx #1\expandafter \@firstoftwo
 \else \expandafter \@secondoftwo
 \fi
}%
\providecommand \natexlab [1]{#1}%
\providecommand \enquote  [1]{``#1''}%
\providecommand \bibnamefont  [1]{#1}%
\providecommand \bibfnamefont [1]{#1}%
\providecommand \citenamefont [1]{#1}%
\providecommand \href@noop [0]{\@secondoftwo}%
\providecommand \href [0]{\begingroup \@sanitize@url \@href}%
\providecommand \@href[1]{\@@startlink{#1}\@@href}%
\providecommand \@@href[1]{\endgroup#1\@@endlink}%
\providecommand \@sanitize@url [0]{\catcode `\\12\catcode `\$12\catcode
  `\&12\catcode `\#12\catcode `\^12\catcode `\_12\catcode `\%12\relax}%
\providecommand \@@startlink[1]{}%
\providecommand \@@endlink[0]{}%
\providecommand \url  [0]{\begingroup\@sanitize@url \@url }%
\providecommand \@url [1]{\endgroup\@href {#1}{\urlprefix }}%
\providecommand \urlprefix  [0]{URL }%
\providecommand \Eprint [0]{\href }%
\providecommand \doibase [0]{http://dx.doi.org/}%
\providecommand \selectlanguage [0]{\@gobble}%
\providecommand \bibinfo  [0]{\@secondoftwo}%
\providecommand \bibfield  [0]{\@secondoftwo}%
\providecommand \translation [1]{[#1]}%
\providecommand \BibitemOpen [0]{}%
\providecommand \bibitemStop [0]{}%
\providecommand \bibitemNoStop [0]{.\EOS\space}%
\providecommand \EOS [0]{\spacefactor3000\relax}%
\providecommand \BibitemShut  [1]{\csname bibitem#1\endcsname}%
\let\auto@bib@innerbib\@empty
\bibitem [{\citenamefont {Fortunato}(2010)}]{fortunato2010community}%
  \BibitemOpen
  \bibfield  {author} {\bibinfo {author} {\bibfnamefont {S.}~\bibnamefont
  {Fortunato}},\ }\href@noop {} {\bibfield  {journal} {\bibinfo  {journal}
  {Physics Reports}\ }\textbf {\bibinfo {volume} {486}},\ \bibinfo {pages} {75}
  (\bibinfo {year} {2010})}\BibitemShut {NoStop}%
\bibitem [{\citenamefont {Yang}\ and\ \citenamefont
  {Leskovec}(2012)}]{yang2012defining}%
  \BibitemOpen
  \bibfield  {author} {\bibinfo {author} {\bibfnamefont {J.}~\bibnamefont
  {Yang}}\ and\ \bibinfo {author} {\bibfnamefont {J.}~\bibnamefont
  {Leskovec}},\ }in\ \href@noop {} {\emph {\bibinfo {booktitle} {Proceedings of
  the ACM SIGKDD Workshop on Mining Data Semantics}}}\ (\bibinfo {organization}
  {ACM},\ \bibinfo {year} {2012})\ p.~\bibinfo {pages} {3}\BibitemShut
  {NoStop}%
\bibitem [{\citenamefont {Girvan}\ and\ \citenamefont
  {Newman}(2002)}]{girvan2002community}%
  \BibitemOpen
  \bibfield  {author} {\bibinfo {author} {\bibfnamefont {M.}~\bibnamefont
  {Girvan}}\ and\ \bibinfo {author} {\bibfnamefont {M.}~\bibnamefont
  {Newman}},\ }\href@noop {} {\bibfield  {journal} {\bibinfo  {journal}
  {Proceedings of the National Academy of Sciences}\ }\textbf {\bibinfo
  {volume} {99}},\ \bibinfo {pages} {7821} (\bibinfo {year}
  {2002})}\BibitemShut {NoStop}%
\bibitem [{\citenamefont {Fortunato}\ and\ \citenamefont
  {Barthelemy}(2007)}]{fortunato2007resolution}%
  \BibitemOpen
  \bibfield  {author} {\bibinfo {author} {\bibfnamefont {S.}~\bibnamefont
  {Fortunato}}\ and\ \bibinfo {author} {\bibfnamefont {M.}~\bibnamefont
  {Barthelemy}},\ }\href@noop {} {\bibfield  {journal} {\bibinfo  {journal}
  {Proceedings of the National Academy of Sciences}\ }\textbf {\bibinfo
  {volume} {104}},\ \bibinfo {pages} {36} (\bibinfo {year} {2007})}\BibitemShut
  {NoStop}%
\bibitem [{\citenamefont {Arenas}\ \emph {et~al.}(2008)\citenamefont {Arenas},
  \citenamefont {Fernandez},\ and\ \citenamefont {Gomez}}]{arenas2008analysis}%
  \BibitemOpen
  \bibfield  {author} {\bibinfo {author} {\bibfnamefont {A.}~\bibnamefont
  {Arenas}}, \bibinfo {author} {\bibfnamefont {A.}~\bibnamefont {Fernandez}}, \
  and\ \bibinfo {author} {\bibfnamefont {S.}~\bibnamefont {Gomez}},\
  }\href@noop {} {\bibfield  {journal} {\bibinfo  {journal} {New Journal of
  Physics}\ }\textbf {\bibinfo {volume} {10}},\ \bibinfo {pages} {053039}
  (\bibinfo {year} {2008})}\BibitemShut {NoStop}%
\bibitem [{\citenamefont {Reichardt}\ and\ \citenamefont
  {Bornholdt}(2006)}]{reichardt2006statistical}%
  \BibitemOpen
  \bibfield  {author} {\bibinfo {author} {\bibfnamefont {J.}~\bibnamefont
  {Reichardt}}\ and\ \bibinfo {author} {\bibfnamefont {S.}~\bibnamefont
  {Bornholdt}},\ }\href@noop {} {\bibfield  {journal} {\bibinfo  {journal}
  {Physical Review E}\ }\textbf {\bibinfo {volume} {74}},\ \bibinfo {pages}
  {016110} (\bibinfo {year} {2006})}\BibitemShut {NoStop}%
\bibitem [{\citenamefont {Aldecoa}\ and\ \citenamefont
  {Mar{\'\i}n}(2013)}]{Aldecoa:2013fk}%
  \BibitemOpen
  \bibfield  {author} {\bibinfo {author} {\bibfnamefont {R.}~\bibnamefont
  {Aldecoa}}\ and\ \bibinfo {author} {\bibfnamefont {I.}~\bibnamefont
  {Mar{\'\i}n}},\ }\href {http://dx.doi.org/10.1038/srep01060} {\bibfield
  {journal} {\bibinfo  {journal} {Sci. Rep.}\ }\textbf {\bibinfo {volume} {3}}
  (\bibinfo {year} {2013})}\BibitemShut {NoStop}%
\bibitem [{\citenamefont {Lancichinetti}\ \emph {et~al.}(2008)\citenamefont
  {Lancichinetti}, \citenamefont {Fortunato},\ and\ \citenamefont
  {Radicchi}}]{lancichinetti2008benchmark}%
  \BibitemOpen
  \bibfield  {author} {\bibinfo {author} {\bibfnamefont {A.}~\bibnamefont
  {Lancichinetti}}, \bibinfo {author} {\bibfnamefont {S.}~\bibnamefont
  {Fortunato}}, \ and\ \bibinfo {author} {\bibfnamefont {F.}~\bibnamefont
  {Radicchi}},\ }\href@noop {} {\bibfield  {journal} {\bibinfo  {journal}
  {Physical Review E}\ }\textbf {\bibinfo {volume} {78}},\ \bibinfo {pages}
  {046110} (\bibinfo {year} {2008})}\BibitemShut {NoStop}%
\bibitem [{\citenamefont {Lancichinetti}\ \emph {et~al.}(2009)\citenamefont
  {Lancichinetti}, \citenamefont {Fortunato},\ and\ \citenamefont
  {Kert{\'e}sz}}]{lancichinetti2009detecting}%
  \BibitemOpen
  \bibfield  {author} {\bibinfo {author} {\bibfnamefont {A.}~\bibnamefont
  {Lancichinetti}}, \bibinfo {author} {\bibfnamefont {S.}~\bibnamefont
  {Fortunato}}, \ and\ \bibinfo {author} {\bibfnamefont {J.}~\bibnamefont
  {Kert{\'e}sz}},\ }\href@noop {} {\bibfield  {journal} {\bibinfo  {journal}
  {New Journal of Physics}\ }\textbf {\bibinfo {volume} {11}},\ \bibinfo
  {pages} {033015} (\bibinfo {year} {2009})}\BibitemShut {NoStop}%
\bibitem [{\citenamefont {Orman}\ \emph {et~al.}(2012)\citenamefont {Orman},
  \citenamefont {Labatut},\ and\ \citenamefont
  {Cherifi}}]{orman2012comparative}%
  \BibitemOpen
  \bibfield  {author} {\bibinfo {author} {\bibfnamefont {G.~K.}\ \bibnamefont
  {Orman}}, \bibinfo {author} {\bibfnamefont {V.}~\bibnamefont {Labatut}}, \
  and\ \bibinfo {author} {\bibfnamefont {H.}~\bibnamefont {Cherifi}},\
  }\href@noop {} {\bibfield  {journal} {\bibinfo  {journal} {Journal of
  Statistical Mechanics: Theory and Experiment}\ }\textbf {\bibinfo {volume}
  {2012}},\ \bibinfo {pages} {P08001} (\bibinfo {year} {2012})}\BibitemShut
  {NoStop}%
\bibitem [{\citenamefont {Labatut}(2013)}]{labatut2013generalized}%
  \BibitemOpen
  \bibfield  {author} {\bibinfo {author} {\bibfnamefont {V.}~\bibnamefont
  {Labatut}},\ }\href@noop {} {\bibfield  {journal} {\bibinfo  {journal} {arXiv
  preprint arXiv:1303.5441}\ } (\bibinfo {year} {2013})}\BibitemShut {NoStop}%
\bibitem [{\citenamefont {Zachary}(1977)}]{zachary1977information}%
  \BibitemOpen
  \bibfield  {author} {\bibinfo {author} {\bibfnamefont {W.}~\bibnamefont
  {Zachary}},\ }\href@noop {} {\bibfield  {journal} {\bibinfo  {journal}
  {Journal of anthropological research}\ ,\ \bibinfo {pages} {452}} (\bibinfo
  {year} {1977})}\BibitemShut {NoStop}%
\bibitem [{\citenamefont {Lusseau}(2003)}]{lusseau2003emergent}%
  \BibitemOpen
  \bibfield  {author} {\bibinfo {author} {\bibfnamefont {D.}~\bibnamefont
  {Lusseau}},\ }\href@noop {} {\bibfield  {journal} {\bibinfo  {journal}
  {Proceedings of the Royal Society of London. Series B: Biological Sciences}\
  }\textbf {\bibinfo {volume} {270}},\ \bibinfo {pages} {S186} (\bibinfo {year}
  {2003})}\BibitemShut {NoStop}%
\bibitem [{\citenamefont {Leskovec}\ \emph {et~al.}(2010)\citenamefont
  {Leskovec}, \citenamefont {Lang},\ and\ \citenamefont
  {Mahoney}}]{leskovec2010empirical}%
  \BibitemOpen
  \bibfield  {author} {\bibinfo {author} {\bibfnamefont {J.}~\bibnamefont
  {Leskovec}}, \bibinfo {author} {\bibfnamefont {K.}~\bibnamefont {Lang}}, \
  and\ \bibinfo {author} {\bibfnamefont {M.}~\bibnamefont {Mahoney}},\ }in\
  \href@noop {} {\emph {\bibinfo {booktitle} {Proceedings of the 19th
  international conference on World wide web}}}\ (\bibinfo {organization}
  {ACM},\ \bibinfo {year} {2010})\ pp.\ \bibinfo {pages} {631--640}\BibitemShut
  {NoStop}%
\bibitem [{\citenamefont {Cazabet}\ \emph {et~al.}(2012)\citenamefont
  {Cazabet}, \citenamefont {Leguistin},\ and\ \citenamefont
  {Amblard}}]{cazabet2012automated}%
  \BibitemOpen
  \bibfield  {author} {\bibinfo {author} {\bibfnamefont {R.}~\bibnamefont
  {Cazabet}}, \bibinfo {author} {\bibfnamefont {M.}~\bibnamefont {Leguistin}},
  \ and\ \bibinfo {author} {\bibfnamefont {F.}~\bibnamefont {Amblard}},\ }in\
  \href@noop {} {\emph {\bibinfo {booktitle} {Proceedings of the 4th
  International Workshop on Web Intelligence \& Communities}}}\ (\bibinfo
  {organization} {ACM},\ \bibinfo {year} {2012})\ pp.\ \bibinfo {pages}
  {1--6}\BibitemShut {NoStop}%
\bibitem [{\citenamefont {Hric}\ \emph {et~al.}(2014)\citenamefont {Hric},
  \citenamefont {Darst},\ and\ \citenamefont {Fortunato}}]{hric2014community}%
  \BibitemOpen
  \bibfield  {author} {\bibinfo {author} {\bibfnamefont {D.}~\bibnamefont
  {Hric}}, \bibinfo {author} {\bibfnamefont {R.~K.}\ \bibnamefont {Darst}}, \
  and\ \bibinfo {author} {\bibfnamefont {S.}~\bibnamefont {Fortunato}},\
  }\href@noop {} {\bibfield  {journal} {\bibinfo  {journal} {arXiv preprint
  arXiv:1406.0146}\ } (\bibinfo {year} {2014})}\BibitemShut {NoStop}%
\bibitem [{\citenamefont {Leskovec}\ \emph {et~al.}(2009)\citenamefont
  {Leskovec}, \citenamefont {Lang}, \citenamefont {Dasgupta},\ and\
  \citenamefont {Mahoney}}]{leskovec2009community}%
  \BibitemOpen
  \bibfield  {author} {\bibinfo {author} {\bibfnamefont {J.}~\bibnamefont
  {Leskovec}}, \bibinfo {author} {\bibfnamefont {K.}~\bibnamefont {Lang}},
  \bibinfo {author} {\bibfnamefont {A.}~\bibnamefont {Dasgupta}}, \ and\
  \bibinfo {author} {\bibfnamefont {M.}~\bibnamefont {Mahoney}},\ }\href@noop
  {} {\bibfield  {journal} {\bibinfo  {journal} {Internet Mathematics}\
  }\textbf {\bibinfo {volume} {6}},\ \bibinfo {pages} {29} (\bibinfo {year}
  {2009})}\BibitemShut {NoStop}%
\bibitem [{Note1()}]{Note1}%
  \BibitemOpen
  \bibinfo {note} {Network available at the following address: \protect \url
  {http://dx.doi.org/10.5281/zenodo.12453"}}\BibitemShut {NoStop}%
\bibitem [{\citenamefont {Gaume}\ \emph {et~al.}(2008)\citenamefont {Gaume},
  \citenamefont {Duvignau}, \citenamefont {Pr{\'e}vot},\ and\ \citenamefont
  {Desalle}}]{gaume2008toward}%
  \BibitemOpen
  \bibfield  {author} {\bibinfo {author} {\bibfnamefont {B.}~\bibnamefont
  {Gaume}}, \bibinfo {author} {\bibfnamefont {K.}~\bibnamefont {Duvignau}},
  \bibinfo {author} {\bibfnamefont {L.}~\bibnamefont {Pr{\'e}vot}}, \ and\
  \bibinfo {author} {\bibfnamefont {Y.}~\bibnamefont {Desalle}},\ }in\
  \href@noop {} {\emph {\bibinfo {booktitle} {Proceedings of the workshop on
  Cognitive Aspects of the Lexicon}}}\ (\bibinfo {organization} {Association
  for Computational Linguistics},\ \bibinfo {year} {2008})\ pp.\ \bibinfo
  {pages} {86--93}\BibitemShut {NoStop}%
\bibitem [{\citenamefont {Newman}(2006)}]{newman2006finding}%
  \BibitemOpen
  \bibfield  {author} {\bibinfo {author} {\bibfnamefont {M.}~\bibnamefont
  {Newman}},\ }\href@noop {} {\bibfield  {journal} {\bibinfo  {journal}
  {Physical review E}\ }\textbf {\bibinfo {volume} {74}},\ \bibinfo {pages}
  {036104} (\bibinfo {year} {2006})}\BibitemShut {NoStop}%
\bibitem [{\citenamefont {Csardi}\ and\ \citenamefont
  {Nepusz}(2006)}]{Csardi:2006uq}%
  \BibitemOpen
  \bibfield  {author} {\bibinfo {author} {\bibfnamefont {G.}~\bibnamefont
  {Csardi}}\ and\ \bibinfo {author} {\bibfnamefont {T.}~\bibnamefont
  {Nepusz}},\ }\href {http://igraph.org} {\bibfield  {journal} {\bibinfo
  {journal} {InterJournal}\ }\textbf {\bibinfo {volume} {Complex Systems}},\
  \bibinfo {pages} {1695} (\bibinfo {year} {2006})}\BibitemShut {NoStop}%
\bibitem [{\citenamefont {Clauset}\ \emph {et~al.}(2004)\citenamefont
  {Clauset}, \citenamefont {Newman},\ and\ \citenamefont
  {Moore}}]{clauset2004finding}%
  \BibitemOpen
  \bibfield  {author} {\bibinfo {author} {\bibfnamefont {A.}~\bibnamefont
  {Clauset}}, \bibinfo {author} {\bibfnamefont {M.}~\bibnamefont {Newman}}, \
  and\ \bibinfo {author} {\bibfnamefont {C.}~\bibnamefont {Moore}},\
  }\href@noop {} {\bibfield  {journal} {\bibinfo  {journal} {Physical review
  E}\ }\textbf {\bibinfo {volume} {70}},\ \bibinfo {pages} {066111} (\bibinfo
  {year} {2004})}\BibitemShut {NoStop}%
\bibitem [{\citenamefont {Pons}\ and\ \citenamefont
  {Latapy}(2005)}]{pons2005computing}%
  \BibitemOpen
  \bibfield  {author} {\bibinfo {author} {\bibfnamefont {P.}~\bibnamefont
  {Pons}}\ and\ \bibinfo {author} {\bibfnamefont {M.}~\bibnamefont {Latapy}},\
  }\href@noop {} {\bibfield  {journal} {\bibinfo  {journal} {Computer and
  Information Sciences-ISCIS 2005}\ ,\ \bibinfo {pages} {284}} (\bibinfo {year}
  {2005})}\BibitemShut {NoStop}%
\bibitem [{\citenamefont {Rosvall}\ and\ \citenamefont
  {Bergstrom}(2011)}]{rosvall2011multilevel}%
  \BibitemOpen
  \bibfield  {author} {\bibinfo {author} {\bibfnamefont {M.}~\bibnamefont
  {Rosvall}}\ and\ \bibinfo {author} {\bibfnamefont {C.~T.}\ \bibnamefont
  {Bergstrom}},\ }\href@noop {} {\bibfield  {journal} {\bibinfo  {journal}
  {PloS one}\ }\textbf {\bibinfo {volume} {6}},\ \bibinfo {pages} {e18209}
  (\bibinfo {year} {2011})}\BibitemShut {NoStop}%
\bibitem [{\citenamefont {Blondel}\ \emph {et~al.}(2008)\citenamefont
  {Blondel}, \citenamefont {Guillaume}, \citenamefont {Lambiotte},\ and\
  \citenamefont {Lefebvre}}]{blondel2008fast}%
  \BibitemOpen
  \bibfield  {author} {\bibinfo {author} {\bibfnamefont {V.}~\bibnamefont
  {Blondel}}, \bibinfo {author} {\bibfnamefont {J.}~\bibnamefont {Guillaume}},
  \bibinfo {author} {\bibfnamefont {R.}~\bibnamefont {Lambiotte}}, \ and\
  \bibinfo {author} {\bibfnamefont {E.}~\bibnamefont {Lefebvre}},\ }\href@noop
  {} {\bibfield  {journal} {\bibinfo  {journal} {Journal of Statistical
  Mechanics: Theory and Experiment}\ }\textbf {\bibinfo {volume} {2008}},\
  \bibinfo {pages} {P10008} (\bibinfo {year} {2008})}\BibitemShut {NoStop}%
\bibitem [{\citenamefont {Aynaud}\ and\ \citenamefont
  {Guillaume}(2010)}]{aynaud2010static}%
  \BibitemOpen
  \bibfield  {author} {\bibinfo {author} {\bibfnamefont {T.}~\bibnamefont
  {Aynaud}}\ and\ \bibinfo {author} {\bibfnamefont {J.}~\bibnamefont
  {Guillaume}},\ }in\ \href@noop {} {\emph {\bibinfo {booktitle} {Modeling and
  Optimization in Mobile, Ad Hoc and Wireless Networks (WiOpt), 2010
  Proceedings of the 8th International Symposium on}}}\ (\bibinfo
  {organization} {IEEE},\ \bibinfo {year} {2010})\ pp.\ \bibinfo {pages}
  {513--519}\BibitemShut {NoStop}%
\bibitem [{\citenamefont {Nicosia}\ \emph {et~al.}(2009)\citenamefont
  {Nicosia}, \citenamefont {Mangioni}, \citenamefont {Carchiolo},\ and\
  \citenamefont {Malgeri}}]{nicosia2009extending}%
  \BibitemOpen
  \bibfield  {author} {\bibinfo {author} {\bibfnamefont {V.}~\bibnamefont
  {Nicosia}}, \bibinfo {author} {\bibfnamefont {G.}~\bibnamefont {Mangioni}},
  \bibinfo {author} {\bibfnamefont {V.}~\bibnamefont {Carchiolo}}, \ and\
  \bibinfo {author} {\bibfnamefont {M.}~\bibnamefont {Malgeri}},\ }\href@noop
  {} {\bibfield  {journal} {\bibinfo  {journal} {Journal of Statistical
  Mechanics: Theory and Experiment}\ }\textbf {\bibinfo {volume} {2009}},\
  \bibinfo {pages} {P03024} (\bibinfo {year} {2009})}\BibitemShut {NoStop}%
\end{thebibliography}%

\end{document}